\documentclass[twocolumn]{emulateapj}
\usepackage{amsmath}
\usepackage{natbib}
\usepackage{color}
\usepackage{graphics}
\usepackage{enumerate}
\usepackage[bookmarks,bookmarksopen,colorlinks,linkcolor={blue},citecolor={blue},urlcolor={blue}]{hyperref}

\shortauthors{Cloutier et al.}
\shorttitle{Could Jupiter or Saturn have ejected a fifth giant planet?}

\begin{document}

\title{Could Jupiter or Saturn have ejected a fifth giant planet?}
  
\author{Ryan Cloutier\altaffilmark{1,2}}
\author{Daniel Tamayo\altaffilmark{2,3}}
\author{Diana Valencia\altaffilmark{2,1}}
\altaffiltext{1}{Dept. of Astronomy \& Astrophysics University 
of Toronto, 50 St. George Street, Toronto, Ontario, Canada, 
M5S 3H4}
\altaffiltext{2}{Centre for Planetary Sciences, 
University of Toronto, Department of Physical \& Environmental 
Sciences, 1265 Military Trail, Toronto, Ontario, Canada, 
M1C 1A4}
\altaffiltext{3}{Canadian Institute for Theoretical Astrophysics, 
60 St. George Street, Toronto, Ontario, Canada, M5S 3H8}

\begin{abstract}
Models of the dynamical evolution of the early solar system following the 
dispersal of the gaseous protoplanetary disk have been widely 
successful in reconstructing the current orbital configuration of the 
giant planets. Statistically, some of the most successful dynamical evolution 
simulations have initially included a hypothetical fifth giant planet, of ice 
giant mass, 
which gets ejected by a gas giant during the early solar system's 
proposed instability 
phase. We investigate the likelihood of an ice giant ejection event by either 
Jupiter or Saturn through constraints imposed by the current orbits of their 
wide-separation regular satellites 
Callisto and Iapetus respectively. 
We show that planetary encounters that are sufficient to 
eject an ice giant, often provide excessive perturbations to the orbits 
of Callisto and Iapetus making it difficult to reconcile a 
planet ejection event with the current orbit of either satellite. 
Quantitatively, we compute the likelihood of reconciling a regular Jovian satellite 
orbit with the current orbit of Callisto following an ice giant ejection by Jupiter 
of $\sim 42$\% and conclude that such a large likelihood supports the hypothesis 
of a fifth giant planet's existence. 
A similar calculation for Iapetus reveals 
that it is much more difficult for Saturn to have ejected an ice giant and reconcile a 
Kronian satellite orbit with that of Iapetus (likelihood $\sim 1$\%), although uncertainties 
regarding the formation of Iapetus, on its unusual orbit, complicates the interpretation of 
this result.
\end{abstract}


\section{Introduction}
Various solar system formation models argue that the giant planets 
underwent planetesimal driven migration 
\citep[e.g.][]{fernandez84,malhotra95,hahn99,tsiganis05} 
at early times ($\lesssim 1$ Gyr) following a dynamical instability.
The \emph{Nice model}, 
originally presented by \cite{gomes05}, \cite{morbidelli05},
and \cite{tsiganis05}, with subsequent 
variants under the same name, 
has been the most successful in reproducing 
the settling of the four giant planets into their present orbital 
configuration \citep{tsiganis05,morbidelli07,levison11}, 
the Late Heavy Bombardment 
at $\sim 700$ Myr \citep{gomes05}, the capture of Jupiter's 
Trojan asteroids \citep{morbidelli05}, 
the capture of gas giant irregular satellites \citep{nesvorny07}, 
as well as the structure of the Kuiper belt 
\citep{levison08} 
and how its dynamical evolution led to the contamination 
of the outer asteroid belt by primitive trans-Neptunian objects 
\citep{levison09}.

The precise nature of giant planet migration in the early solar system 
remains uncertain due to our lack of knowledge regarding each body's initial 
conditions following their formation out of the solar nebula and the 
chaotic nature of the migration process. 
However, \cite{morbidelli09} argued that smooth 
divergent migration of the gas giants is unable to 
sufficiently excite their 
orbital eccentricities and inclinations to their observed values. 
Additionally, \cite{brasser09} 
showed that such migration from an initial 
resonant configuration following the dispersal of the gaseous disk  
leads to excessive orbital 
eccentricities in the previously formed terrestrial bodies 
via sweeping secular resonances. A proposed solution, 
known as the \emph{jumping-Jupiter scenario} \citep{brasser09}, invokes 
close encounters between the gas giants and an ice giant (IG) resulting 
in the step-wise migration of Jupiter and Saturn from their 
initial mean-motion resonance. 
This can sufficiently excite giant planet eccentricities and inclinations 
whilst jumping over the problematic secular frequencies of the terrestrial 
planets. In addition, the \emph{jumping-Jupiter scenario} 
does not disrupt the asteroid belt's observed morphology \citep{morbidelli10}.
   
A statistical study by \cite{nesvorny11} 
of the dynamical evolution of the 
solar system during such a phase of frequent planetary encounters showed that
the likelihood of reconstructing the current orbital configuration of the 
four giant planets is increased when a fifth giant planet of approximately 
Uranian mass is included in the early solar system. 
The instability, which gives rise to multiple planetary 
encounters, results in the ejection  
of the hypothetical fifth giant planet, reconstructing the outer solar 
system whilst preserving the orbits of the inner terrestrial bodies 
over long timescales \citep{batygin12,nesvorny12}. 
Such planet scattering events \citep{rasio96,weidenschilling96}, 
applicable to any multi-body system, 
provide a potential explanation for the existence of the recently 
detected ``free-floating'' planets \citep[e.g.][]{delorme12,liu13,luhman14}.

In addition to reconstructing the current orbital configuration of both 
large and small bodies in the solar system, 
models attempting to achieve a full description of the solar system's 
early dynamical evolution must require the survival of the giant planets' 
regular satellites \citep[e.g.][]{deienno14}. 
Regular satellites which are thought to form via 
accretion processes in circumplanetary disks \citep{canup02,mosqueira03}, 
are expected to form on prograde, low-eccentricity orbits that are 
nearly coplanar with the host's equatorial plane and have 
relatively small semimajor axes. 
The current deviations of Callisto and Iapetus' orbits from circular, 
uninclined orbits therefore limit how close an IG could have come to the 
gas giants in the early solar system.

In general, outer satellites, which are less tightly bound to the planet, will
suffer larger perturbations during a close approach with an IG.  In addition, the
orbital eccentricities of these outer moons are only marginally damped through tides
(which could otherwise mask the effects of early encounters).  It is therefore the
outermost regular satellites (Callisto around Jupiter and Iapetus around Saturn)
which provide the most stringent constraints.  

Specifically, \cite{deienno14} investigated whether the close encounters in the
particular simulations of \cite{nesvorny12} (referred to as NM12), 
that best reproduced the giant planets’ orbital architecture, would excessively excite
Callisto's orbit.  Given the interest in additional planets in the early solar
system, we generalize this question to ask how likely it is to retain Callisto 
(Iapetus) at its observed orbit following an ejection of an IG by Jupiter (Saturn)? 
Although in the \emph{jumping-Jupiter scenario} only Jupiter may be responsible 
for ejecting an IG, we include an analysis of close IG encounters with Saturn as we 
are more generally interested in early solar system instability models in which 
either gas giant could undergo close encounters with the solar system's IGs. It also 
permits the direct comparison of the likelihood of retaining a Callisto-like satellite 
orbit to the likelihood of retaining an Iapetus-like satellite orbit following the ejection 
of an IG (see Sect.~\ref{prob}).

In Sect.~\ref{sats} we discuss the relevant properties 
of the satellites of interest, Sect.~\ref{model} summarizes our methods 
of investigation, Sect.~\ref{jupsect} \& \ref{satsect} 
present our study's results, and Sect.~\ref{prob} 
presents our calculation of the likelihood of reconciling satellite orbits 
following planetary encounters with present-day orbits. We conclude with a detailed 
discussion in Sect.~\ref{discuss} and a summary in Sect.~\ref{summ}.

\section{Summary of Satellites: Callisto \& Iapetus} \label{sats}
Callisto is the outermost Galilean satellite, moving on a 
nearly circular ($e = 0.007$) and uninclined ($i \sim 0.28^{\circ}$) 
orbit.  
Callisto's orbital period is $\sim 16.7$ days and is the only Galilean satellite not 
locked in a mean-motion resonance \citep{musotto02}. 
Iapetus' orbit is somewhat more eccentric ($e \sim 0.03$) and circles Saturn every 
$\sim 79$ days. Curiously, Iapetus exhibits a significantly inclined orbit 
possibly due to inclined planetary encounters between 
Saturn and an IG like those expected in the \emph{jumping-Jupiter scenario} 
\citep{nesvorny14}. Current 
satellite orbital elements are summarized in Table \ref{planets}. These data were obtained 
from JPL HORIZONS.\footnote{http://ssd.jpl.nasa.gov/?horizons}

\begin{table}
\centering
\caption{Summary of Giant Planets and Satellites. \label{planets}}
\begin{tabular}{clll}
  \hline
  \hline
  \textbf{Planet} & & \emph{Jupiter} & \emph{Saturn} \\
  \hline
  &$M_p$ ($M_{\odot}$) & $9.54 \times 10^{-4}$ &  $2.86 \times 10^{-4}$ \\
  &$a_p$ (AU) & 5.20 & 9.54 \\
  &$R_p$ (AU) & $4.67 \times 10^{-4}$ & $3.89 \times 10^{-4}$ \\
  & & & \\
  \textbf{Satellite} && \emph{Callisto} & \emph{Iapetus} \\
  \hline
  &$P_s$ (days) & 16.69 & 79.33 \\  
  &$a_s$ ($R_p$) & 26.93 & 61.15 \\
  &$e_s$ & $7.0 \times 10^{-3}$ & $2.83 \times 10^{-2}$ \\
  &$i_s$ ($^{\circ}$) & 0.28 & 7.49 \\
  \hline
\end{tabular}
\tablecomments{Data are from JPL HORIZONS: http://ssd.jpl.nasa.gov/?horizons}
\end{table}

While Callisto does not participate in a MMR like the other Galilean satellites, 
its eccentricity evolution is nevertheless secularly coupled to that of the inner 
moons \citep{greenberg11}. 
The eccentricities of the inner satellites are more 
easily damped by tides than Callisto because of their smaller semimajor axes and as a result 
of the secular coupling, $e_{Callisto}$ is damped on timescales faster than expected for 
an isolated planet-satellite system that is tidally locked. 
We take this into account as described in Sect.~\ref{jupresultssub} following 
numerical simulations of this effect by \cite{deienno14}.
Conversely, the eccentricity damping of Iapetus since the solar system's 
instability phase, has been negligible \citep{castillo07}.

\section{Methods} \label{model}
We model the evolution of an IG heading for a close approach with one of Jupiter or Saturn, 
and investigate 
the encounter's effect on the gas giant's satellites. We first determine viable encounter 
parameters that 
lead to the ejection of the IG. Within this set, we investigate if there are any satellites 
remaining around Jupiter (Saturn) with an orbit consistent with that of Callisto (Iapetus). 

\subsection{Numerical Model: Simulating Planetary Encounters} \label{planetmodel}
We consider a reduced-body subset of the solar system including the Sun,  
a satellite-hosting gas giant planet and an IG.  
The gas giant of mass $M_p$ is initialized on a osculating, circular orbit 
with semimajor axis $a_p$. By approximating the gas giant's orbit as circular, we remove any 
dependence on its orbital phase at the epoch of encounter; $t_{enc}$. 
Because the mass of the ejected IG is not well constrained, 
we select a fiducial value approximately equal to the mass of Uranus 
($M_I=5\times 10^{-5} M_{\odot}$) as was used by NM12.
A system of Keplarian satellites is placed in orbit around the gas giant 
(see Sect.~\ref{methsats} for a detailed description of satellite orbits).

\begin{figure*}
\plottwo{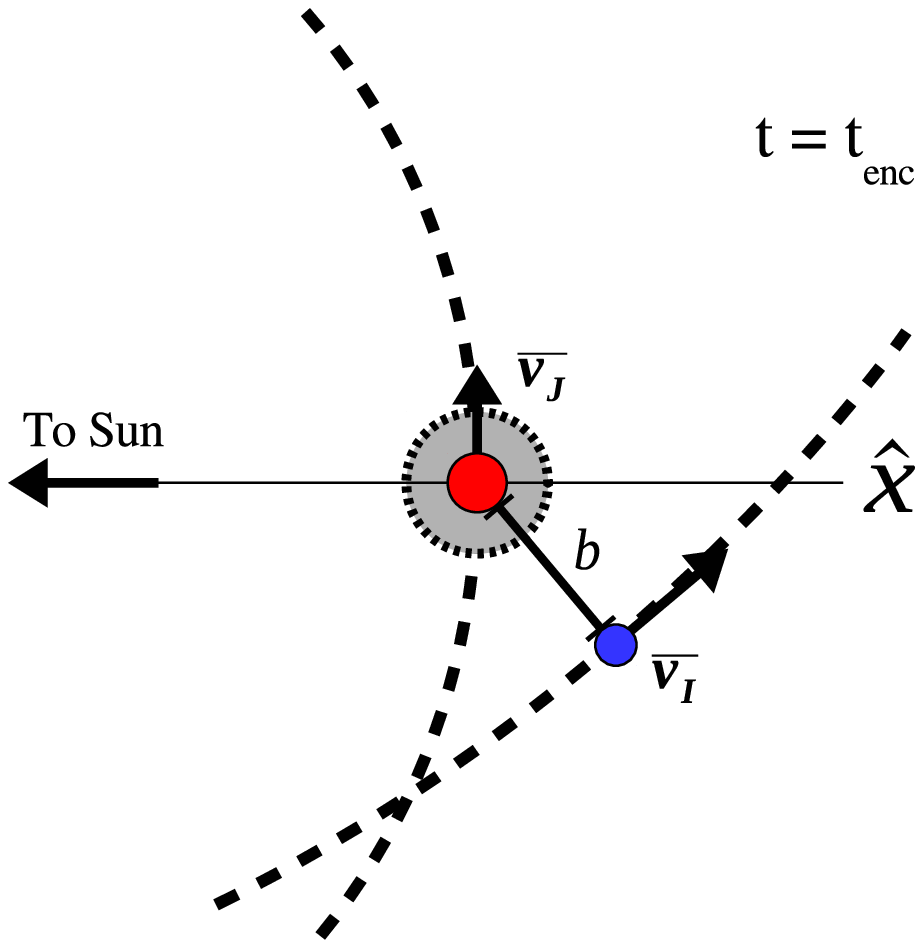}{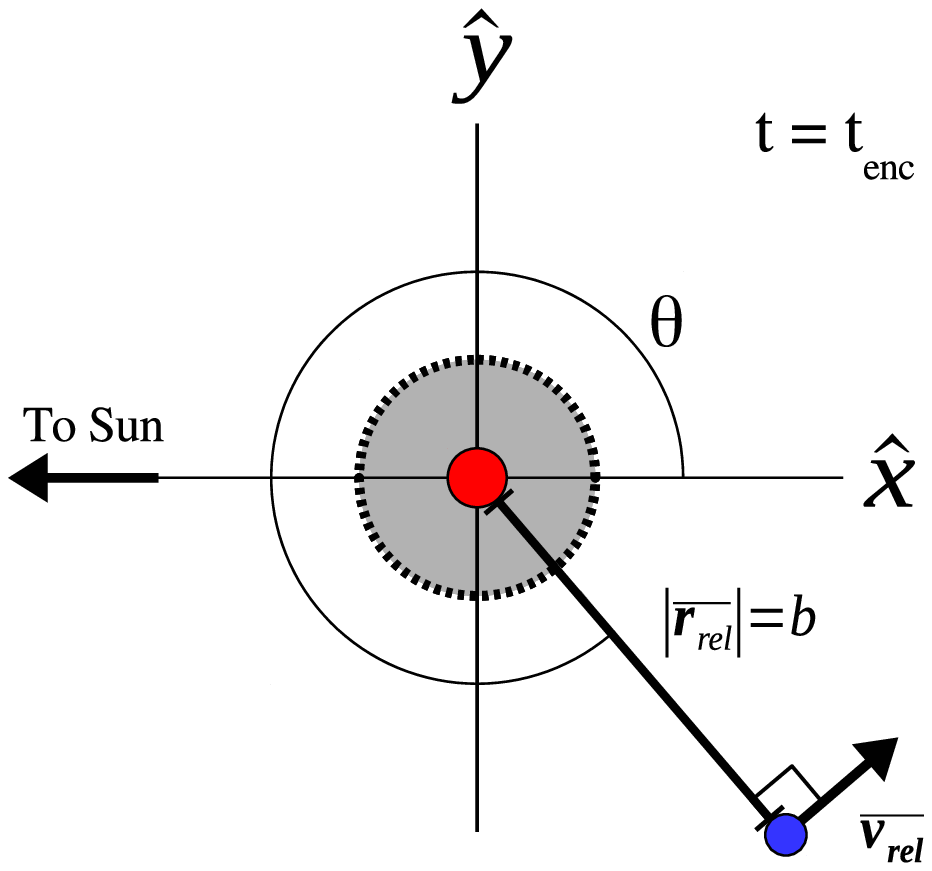}
\caption{Pictorial representations of a close planetary encounter between 
a gas giant (\emph{larger, red planet}) and an unspecified ice giant (IG; 
\emph{smaller, blue planet}) at the epoch of encounter $t_{enc}$. The coplanarity 
of planetary encounters permits analysis solely in the $xy$-plane ($z=0$).  
\emph{Left}: the heliocentric reference frame with the Sun, at a distance of 
the gas giant's semimajor axis, aligned with the $\hat{x}$ axis. The planet's 
instantaneous 
velocity vectors are shown along with their trajectories (\emph{dashed lines}) 
in the vicinity of $t_{enc}$. 
\emph{Right}: the reference frame is centered on the gas giant and is zoomed-in to 
depict the phase angle $\theta$; the angle between the $\hat{x}$ axis 
and the IG's position vector ($\textbf{r}_{rel}$) 
at $t_{enc}$, measured in the counter-clockwise direction. Here, $\theta \approx 310^{\circ}$. 
At $t_{enc}$, the magnitude of the IG's relative 
position vector is equal to the impact parameter $b$ and is orthogonal to the relative 
velocity vector $\textbf{v}_{rel}$. 
In both diagrams, the \emph{dotted grey rings} represent the ring of regular satellites 
in orbit around the gas giant. The scale used here is approximate as these schematics are not 
intended to be exact, but instead, are included  
to aid in the reader's visualization of the experimental setup. \label{schematic}}
\end{figure*}

To limit the number of encounter parameters and computational cost of our 
survey, we assume a coplanar geometry (we present a more detailed discussion of the effect of 
inclined encounters in Sect.~\ref{inclined}). 
Thus, at closest approach the IG's velocity is perpendicular to its separation vector from the 
gas giant and the encounter is fully specified at this time ($t_{enc}$) 
by the impact parameter $b$, the relative 
planet velocity $v_{rel}$, and the phase angle $\theta$ (see Fig. \ref{schematic}). 

We linearly step the impact parameter outward from $b_{min} = 0.02$ AU until
encounters no longer lead to ejections. \cite{deienno14} found that close
encounters with $b < b_{min}$ excessively excite the Galilean satellites; our
estimates in Sect.~\ref{prob} for the likelihood of retaining the observed satellite orbits
following an ejection therefore represent a conservative overestimate due to our
comparatively gentle ejections. However, we note that values smaller than $b_{min}$
are less likely due to the reduced encounter cross-section at small radii. We
estimate the size of this effect by including a reduced simulation sample with 
$b < b_{min}$ and find that our final results (Sect.~\ref{prob}) changed (fractionally) by 
$\lesssim 12$\%. 

To determine which encounter 
parameters are capable of ejecting the IG, we uniformly sample 20,000 parameter 
combinations within $b \in [b_{min},0.1$ AU], $v_{rel} \in [1,5] v_{esc}$, and 
$\theta \in[0,2\pi)$, where $v_{esc} = \sqrt{2G M_p/b}$.  
We then remove any unphysical encounters in which the IG is unbound from the Sun 
prior to the encounter. In the cases of Jupiter 
and Saturn we find $N_{sim}=278$ and $N_{sim}=274$ valid parameter combinations 
respectively.

To simulate an encounter with the aforementioned parameters from Fig. \ref{schematic}, 
we integrate \emph{backwards} in time until the absolute separation of the 
planets $|\textbf{r}_I - \textbf{r}_p| \geq 2$ AU. 
At this separation a 
satellite with a Callisto-like or Iapetus like orbit feels a force from its host which is 
$> 10^{4}$ times greater than that felt from the IG. The positions and 
velocities of all three massive bodies are then used as initial conditions 
for the \emph{forward} simulations which include the satellites. These simulations 
are integrated \emph{forward} in time towards the encounter at  $t=t_{enc}$ and are halted  
at $t=2t_{enc}$. After $2t_{enc}$ the influence of the IG on the satellites is 
again negligible and satellite orbits are no longer perturbed by the IG's influence. 
In order for close encounters with $b<0.1$ AU to be possible, numerous `soft' 
encounters between the gas giants and IG, prior to $t_{enc}$,  
were needed in order to build-up the IG's eccentricity. These 
encounters will supply a small perturbation to regular satellite orbits that, on average, 
increase satellite eccentricity over time (see \cite{nesvorny14} Fig. 4). However, 
the effect of encounters on satellite orbits is strongly dependent on $b$, which 
is much larger than 0.1 AU for `soft' encounters. We are therefore 
only concerned with the strongest (final) encounter which leads to ejection of the IG 
as its effect dominates the final satellite orbits.

Finally, we do not include the perturbations from the satellite host's oblateness.  
This should be a good approximation since the encounter timescale with the IG is much shorter 
than the precession timescales of the satellites due to the non-spherical shape of 
its host planet.

\subsection{Numerical Model: Initializing Satellites} \label{methsats}
In each encounter simulation, we include a ring of Callisto or Iapetus 
analog satellites around the gas giant planet. 
The satellite ring consists of $N=100$ non-interacting 
test particles with azimuthal positions randomly 
sampled from a uniform distribution between $0$ and $2\pi$ 
as to remove any azimuthal dependence at $t_{enc}$. Modelling  
the satellite system as an ensemble of test particles ensure that 
the orbital evolution of the satellites is governed solely by gravitational 
interactions with the host planet and perturbations from the IG and the Sun.

The semimajor axes of 
the satellites $a_s$ are initialized to $\pm 1$\% of the current orbital radius 
of Callisto or Iapetus. This fractional deviation is chosen to be on the order  
of the observed satellite eccentricities $e_s$ (see Table \ref{planets}).
Because changes in $a_s$ and $e_s$ are related through changes in the 
satellites' angular momentum, it is unlikely that larger shifts 
in $a_s$ would be reconcilable with the observed $e_s$ assuming 
that each satellite formed out of a circumplanetary disk  
on a circular orbit \citep{canup02,mosqueira03}.  
Satellites are initialized  
on nearly coplanar and circular orbits ($i \leq 0.1^{\circ}, e \leq 10^{-5}$) 
as expected from circumplanetary formation scenarios.
Satellite particles are not allocated physical sizes as we do not account for 
particle collisions in our simulations. 

\subsection{Numerical Code}
We performed our simulations using the  
\texttt{REBOUND} N-body numerical code \citep{rein12}. 
We employ the \texttt{IAS15} integrator 
(Integrator and Adaptive Step-size control, 15th order;
\citealt{rein15}) whose adaptive timestepping ensures 
optimal resolution of the short satellite orbital periods 
and rapid encounter timescales.\footnote{A short video from \texttt{REBOUND} 
depicting a close planetary encounter with a ring of regular satellites can 
be found \href{http://www.astro.utoronto.ca/~cloutier/rebound_encounter.mp4}{here}.} 

\section{Results of Ice Giant Ejections by Jupiter} \label{jupsect}
Here we focus on simulations in which Jupiter is the satellite-hosting 
gas giant planet that ejects the hypothetical fifth giant planet from the 
solar system. 

\subsection{Properties of Encounters} \label{jupprop}
\begin{figure}
\epsscale{.9}
\includegraphics[scale=.43]{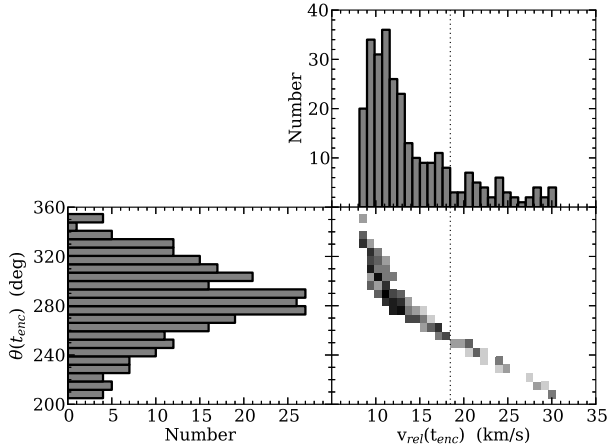}
\caption{Distributions of the IG phase angle ($\theta$; \emph{bottom-left}) and 
relative planet velocity (\emph{top-right}) at $t_{enc}$ for encounters 
between Jupiter and an unspecified IG that resulted in the ejection of the 
latter. Relative 
planet velocities are uniquely determined by the impact parameter $b$ and the  
specified fraction of the escape speed from Jupiter. 
\emph{Bottom-right}: 2D histogram exhibiting the correlation between phase angle 
and relative velocity. 
The darkness of each bin is indicative of the logarithmic 
number of successful ejections. The \emph{vertical dotted line} is indicative of 
the escape velocity from the Sun at the semimajor axis of Jupiter. \label{jupsims}}
\end{figure}

Fig. \ref{jupsims} summarizes the properties of the planetary encounters  
which result in the ejection of the IG planet. 
We find $N_{sim}=278$ such encounters.  
The greatest impact parameter we find capable of ejecting the IG 
is $\approx 0.05$ AU.

All successful ejections involve closest approach 
in the lower hemisphere of the xy-plane in Jupiter's reference frame 
($180^{\circ} < \theta(t_{enc}) < 360^{\circ}$; see Fig. \ref{schematic}).  
As familiar from spacecraft gravity assists, an IG trailing Jupiter at $t_{enc}$ 
will receive a positive velocity kick via their interaction.  
At these phase angles, the relative velocity vector is rotated by 
the encounter such that the IG's inertial speed accelerates.
Depending on $b$ and $v_{rel}$, the encounter can potentially 
boost the IG to escape velocity from the solar system.    
From simple vector diagrams, the maximum increase to the IG velocity
can be shown to occur when $\theta(t_{enc})=270^{\circ}$. Fig. \ref{jupsims} highlights this 
as the distribution of $\theta(t_{enc})$ peaks at $\sim 270^{\circ}$ where the encounter 
geometry is most conducive to ejecting the IG. 

Low $v_{rel}$ trajectories lead more often to ejection since Jupiter can more effectively 
deflect the IG's path. 
We note that there are many encounters in which the velocity of the 
IG with respect to Jupiter at $t_{enc}$ is greater than the escape speed from 
the solar system at Jupiter's distance from the Sun. 
This is due to the IG being accelerated upon approach to the encounter which occurs 
deep in Jupiter's gravitational well. Such 
cases consist of the IG becoming unbound from the solar system prior to the realization 
of the impact parameter at $t_{enc}$ and are thus accelerated to super-escape speeds 
even before its closest approach to Jupiter.

The 2D histogram in Fig. \ref{jupsims} depicts the correlation between the relative 
velocity of the planets at the epoch of encounter and the phase angle in the 
jovian-centric reference frame. 
Encounters which occur with the IG at a smaller heliocentric distance than 
Jupiter ($\theta(t_{enc}) < 270^{\circ}$) require larger relative velocities  
in order to eject the IG. As the IG moves outwards (increasing $\theta$) the  
relative velocities necessary to eject the IG 
decrease because the pull applied by Jupiter following the gravity assist 
is less efficient at decelerating the IG to sub-escape speeds given the IG's 
trajectory.

\subsection{Resulting Jovian Satellite Orbits} \label{jupresultssub}
\begin{figure}
\epsscale{.9}
\includegraphics[scale=.43]{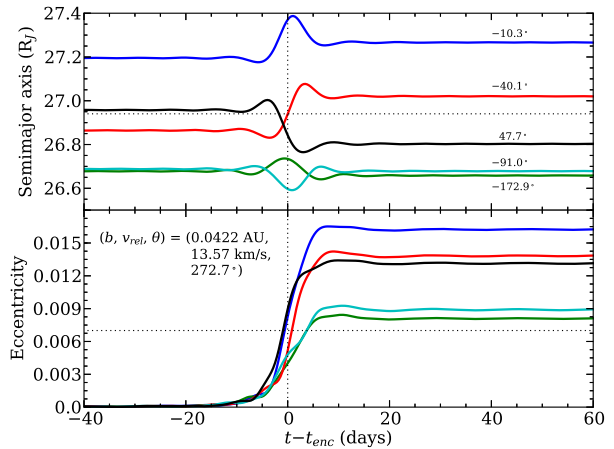}
\caption{Evolution of five sample Jovian satellites as Jupiter undergoes a close encounter with an 
unspecified IG. The encounter parameters at $t_{enc}$: impact parameter $b$, relative planet 
velocity $v_{rel}$, and IG phase angle $\theta$ are shown in the legend of the lower panel. 
Each satellite's azimuthal position relative to $\theta(t_{enc})$ is annotated in the upper panel. 
Vertical and horizontal \emph{dotted lines} indicate the epoch of encounter and the current values of 
Callisto's orbital elements respectively. \label{jupevol}}
\end{figure}

Close planetary encounters between Jupiter and the IG 
can result in significant perturbations to the orbits of the in-situ 
Jovian satellites. The time evolution of $a_s$ and $e_s$ for 
five sample Jovian satellites throughout one 
encounter simulation demonstrates this fact (Fig. \ref{jupevol}).

The five sample satellites which are initialized on nearly circular orbits get kicked to moderately 
eccentric orbits by a planetary encounter with $b \approx 0.042$ AU and $v_{rel} \approx 13.57$ km/s. 
Hence, the characteristic encounter timescale is $b/v_{rel} \sim 5.3$ days or about one third 
of Callisto's orbital period. The orbital perturbation therefore acts roughly as an impulse to 
the satellites making their azimuthal position at $t_{enc}$, a parameter of importance. 
Indeed, in Fig. \ref{jupevol} the satellites 
whose azimuthal positions are increasingly distant from $\theta(t_{enc})$ are on average 
less perturbed than satellites which are close to $\theta(t_{enc})$. 
In addition, the position of the satellite relative to 
$\theta(t_{enc})$ determines the direction of the satellite's radial shift. For example, 
satellites immediately trailing the IG at $t_{enc}$ will be pulled forward in their orbits 
(e.g. \emph{red curve}) increasing $a_s$, whereas satellites immediately 
upstream of the IG at $t_{enc}$ get pulled downward, thus decreasing $a_s$ (e.g. \emph{black curve}). 

\begin{figure}
\epsscale{.9}
\includegraphics[scale=.43]{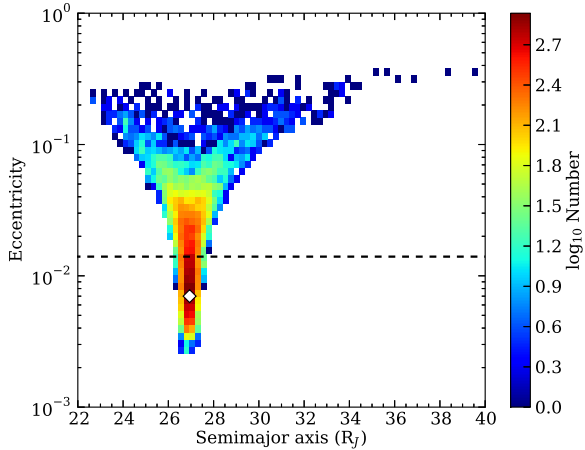}
\caption{2D histogram of the final average semimajor axes and eccentricities of Jovian satellites. 
The colorbar is indicative of the logarithmic 
number of satellites in each bin. The current values of Callisto's 
semimajor axis and eccentricity are depicted on the plot as a \emph{white diamond}. The 
\emph{horizontal dashed line} highlights $2e_{Callisto}$ which marks the boundary between satellites that 
can ($e_s \leq 2e_{Callisto}$) and cannot ($e_s > 2e_{Callisto}$) be reconciled with Callisto's current orbit.  
\label{jupresults}}
\end{figure}

The average final orbital elements of all Jovian satellites are 
depicted in Fig. \ref{jupresults} and show close agreement with the 
results from \cite{deienno14} despite the broader investigation of  
the close encounter parameter space.
Changes in $a_s$ and $e_s$ are coupled through the satellite's orbital angular momentum.  
This effect is highlighted in Fig. \ref{jupresults}, as satellites which are radially 
transferred to either a significantly larger or smaller semimajor axis are those 
which exhibit the largest deviation from a circular orbit. 

For each satellite 
we define a reconcilable orbit to be when the satellite's final average eccentricity 
is $\leq 2e_{Callisto}$ (recall $e_{Callisto} \approx 0.007$).  
The factor of 2 in our definition 
comes from the subsequent eccentricity evolution due to tidal damping of Callisto    
\citep[][c.f. Fig. 7]{deienno14} and allows for Callisto to be excited beyond the present 
$e_{Callisto}$ at $t_{enc}$ and consequently settle into its 
current orbital eccentricity in the 4 Gyrs following the solar system's instability phase. 

For the remainder of the paper we refer to the event of 
a simulated satellite's final orbit being reconcilable with Callisto as 
$RS$ for ``reconcilable satellite''. 
The boundary dividing $RS$ from non-$RS$ satellites is depicted in Fig. \ref{jupresults} 
as a dashed horizontal line. 

Final average $e_s$ values are never $\geq 0.4$, implying that 
no Jovian satellite becomes unbound from Jupiter following the ejection of the IG.  
That is, the vast majority of planetary encounters that are capable of ejecting an IG from the solar 
system are not sufficiently violent to strip Jupiter's regular satellites.  
This favours the possibility of Jupiter being able to eject an IG whilst retaining a regular 
satellite whose orbit is Callisto-like. We estimate the likelihood in Sect.~\ref{prob}.
  
\begin{figure*}
\epsscale{1}
\centering
\includegraphics[scale=.6]{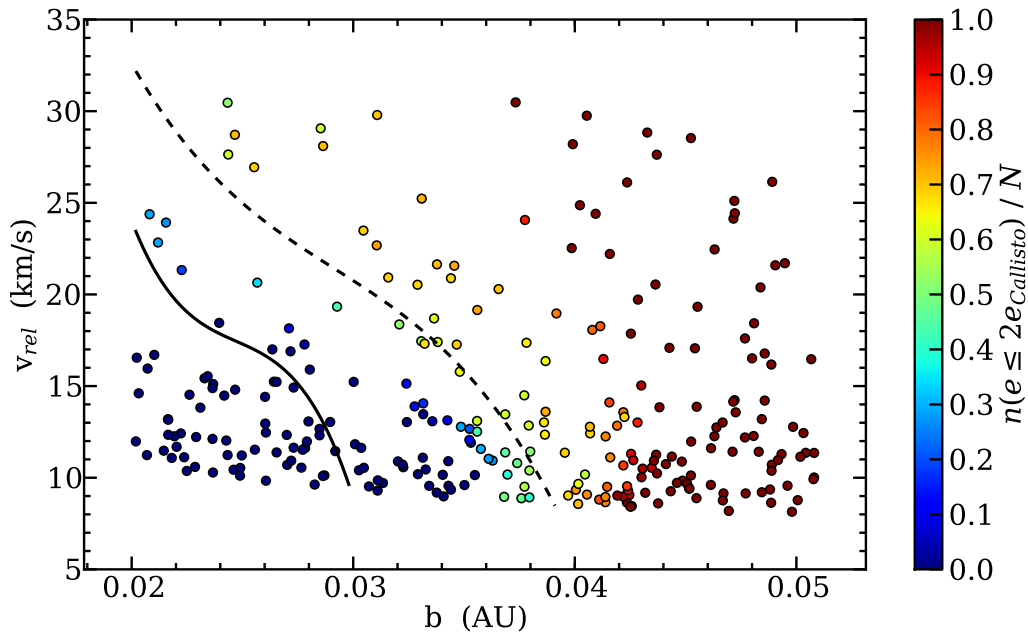}
\caption{The fraction of reconcilable Jovian satellites as a function of the encounter's impact 
parameter and relative planet velocity. The colorbar indicates the fraction of satellites whose 
final average eccentricity is less than or equal to $2e_{Callisto}$. 
Smoothed contours at 10\% (\emph{solid line}) and 50\% (\emph{dashed line}) assist in visualizing 
the different regions of the parameter space. The survival fractions shown here reveal that Jupiter 
has a reasonably large probability of ejecting an IG while retaining a Callisto-like satellite (see 
Sect.~\ref{prob}). \label{finalecall}}
\end{figure*}

While the phase angle $\theta(t_{enc})$ is an important parameter in determining whether 
the IG is ejected, it has little effect on the fraction of perturbed satellites due to 
their uniformly distributed azimuthal positions around Jupiter. 
Therefore, for each $b$ and $v_{rel}$ we can marginalize over $\theta$ to show in Fig. 
\ref{finalecall} the fraction of satellites reconcilable with the orbit 
of Callisto after each encounter. Interpolating 
over the fraction of reconcilable Jovian satellites in ($b$,$v_{rel}$) space, the resulting 
high resolution contours at 10\% and 50\% are fitted with cubic functions and over-plotted 
as solid and dashed curves respectively to aid in visualization of the different regions 
of the parameter space. 

It is clear that as encounters 
become closer, the perturbation to $e_s$ increases and the fraction of 
reconcilable satellites shrinks. Similarly, as the duration of 
encounters becomes longer (smaller $v_{rel}$), the timescale over which the perturbation 
is applied grows, thus increasing the deviation of satellite orbits from circular. 
Hence, forming and maintaining a Jovian satellite with a Callisto-like orbit favours encounters 
which are wide and fast. 

We suggest that researchers simulating solar system formation scenarios can use Fig.
\ref{finalecall} to estimate whether or not a given Jupiter/IG encounter is consistent with
Jupiter's current Galilean satellite orbits.  We note that our requirement that the
IG gets ejected only determines what regions of the plot are populated, hence the
contours are useful guides regardless of whether or not the IG survives the encounter 
(and for
any $\theta$). We caution, however, that these will only be approximate due to our 
assumption of coplanarity. For a given $b$ and $v_{rel}$, inclined encounters will 
increase the fraction of satellites whose eccentricities are reconcilable with 
current orbits. However, they will also tend to overly excite the satellite 
inclinations. We discuss this further in Sect.~\ref{inclined}. For particular simulations 
that lie on the boundary of plausibility in Fig. \ref{finalecall}, 
one could re-simulate the close approaches, including inclined encounters, 
following our setup to more accurately quantify the effect. 

\section{Results of Ice Giant Ejections by Saturn} \label{satsect}
Here we focus on simulations in which Saturn is the satellite-hosting 
gas giant planet that ejects the hypothetical fifth giant planet from the 
solar system. 
\subsection{Properties of Encounters}
\begin{figure}
\epsscale{.9}
\includegraphics[scale=.43]{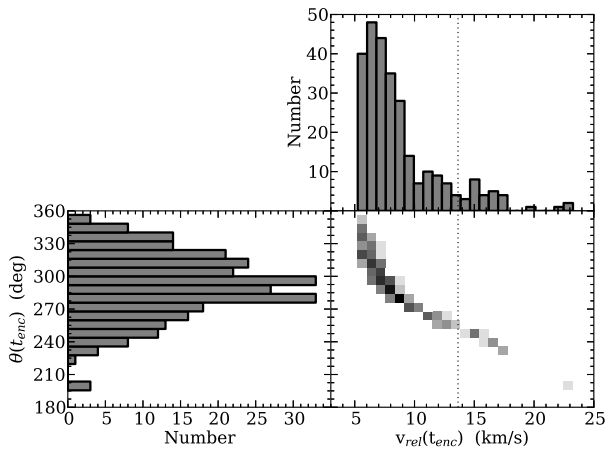}
\caption{Same as Fig. \ref{jupsims} for encounters between Saturn and 
an unspecified IG. The \emph{vertical dotted line} is indicative of 
the escape velocity from the Sun at the semimajor axis of Saturn. \label{satsims}}
\end{figure}

The properties of planetary encounters between Saturn and an IG are summarized 
in Fig. \ref{satsims}. We find $N_{sim} = 274$ simulations which result in the IG being 
ejected by Saturn. The largest impact parameter that is still capable of ejecting the 
IG is nearly identical to the Jupiter/IG case; $\approx 0.05$ AU. 
We perform an additional sampling of encounter parameters with $b>0.05$ AU at 
an increased resolution  
to confirm that the largest impact parameter capable of ejecting the IG is nearly the same in 
both the Jupiter and Saturn cases rather than being a statistical anomaly due to the 
finite sampling of encounter parameters. 
We find that this limit is real and not an artifact of our 
sampling procedure. The physical interpretation of this similarity 
is beyond the scope of this paper.   

Similar trends to those shown in Fig. \ref{jupsims} are observed. 
This should be expected because, modulo the variations in the physical parameters 
$M_p$ and $a_p$, the Saturn/IG encounters investigated 
are fundamentally equivalent to those described in Sect.~\ref{jupprop}.
With the exception of the decreased magnitude of $v_{rel}$ by a factor of 
$\sqrt{M_{Sat}/M_{Jup}}$ on average,  
the set of successful scattered-by-Saturn simulations are statistically identical 
to those in the scattered-by-Jupiter simulations; i.e. the histograms in Figs. \ref{jupsims} 
and \ref{satsims} exhibit the same behaviour. 

\subsection{Resulting Kronian Satellite Orbits}
\begin{figure}
\epsscale{.9}
\includegraphics[scale=.43]{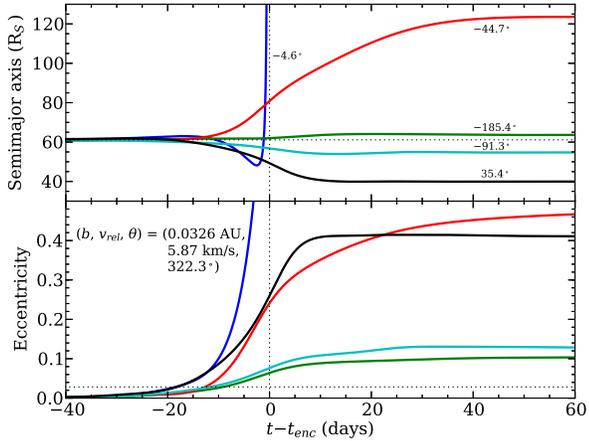}
\caption{Same as Fig. \ref{jupevol} featuring the evolution of five sample Kronian satellites 
as Saturn undergoes a close encounter with an unspecified IG. 
Vertical and horizontal \emph{dotted lines} indicate the epoch of encounter and the current values of 
Iapetus' orbital elements respectively. \label{satevol}}
\end{figure}

\begin{figure}
\epsscale{.9}
\includegraphics[scale=.43]{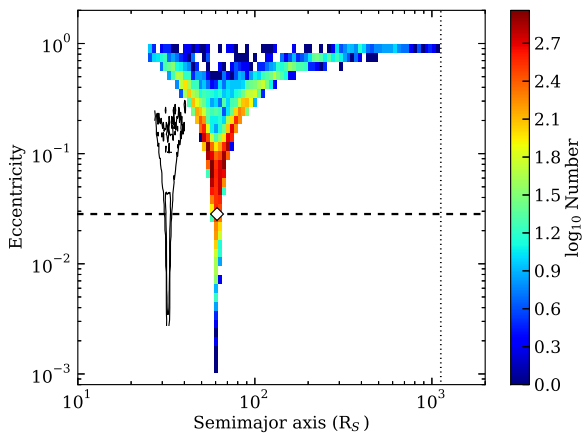}
\caption{Similar to Fig. \ref{jupresults} for the Kronian satellites with the inclusion of 
Saturn's Hill radius ($\sim 1100$ $R_S$; \emph{vertical dotted line}) and contours of 
the 2D histogram shown for Jupiter (Fig. \ref{jupresults}) for direct comparison. 
The \emph{horizontal dashed line} highlights $e_{Iapetus}$ which marks the boundary between 
satellites that can ($e_s \leq e_{Iapetus}$) and cannot ($e_s > e_{Iapetus}$) be reconciled 
with Iapetus' current orbit. Note the logarithmic x-axis. \label{satresults}}
\end{figure}

\begin{figure*}
\centering
\includegraphics[scale=.6]{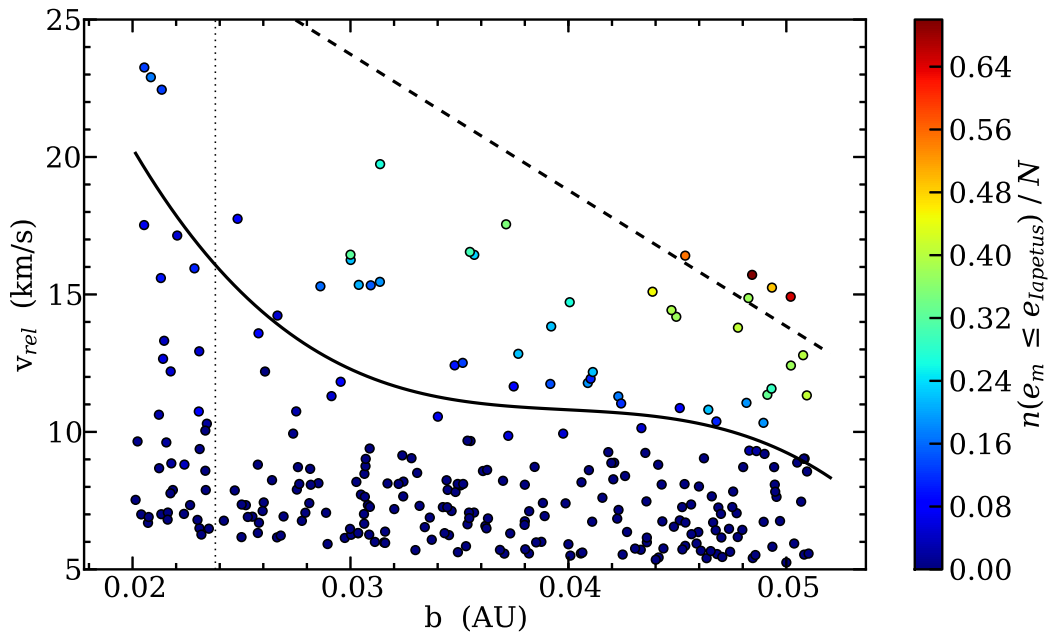}
\caption{Same as Fig. \ref{finalecall} for the Kronian satellites. The colorbar indicates the 
fraction of satellites whose final average $e_s$ is less than or equal to $e_{Iapetus}$. The 
\emph{vertical dotted line} indicates Iapetus's current semimajor axis. The survival 
fractions here reveal that Saturn has a small probability of ejecting an IG while retaining 
an Iapetus-like satellite if we assume that Iapetus formed like a regular satellite 
out of a circumplanetary disk. \label{finaleiap}}
\end{figure*}

Similarly to the case of Jupiter ejecting the IG from the solar system, the Kronian 
satellite orbits will be perturbed as a result of the encounter between Saturn and the IG. 
As a result of the wide separation 
of Iapetus ($\sim 61$ $R_S$) and the similar nature of the encounters between an IG and 
either Jupiter or Saturn (Figs. \ref{jupsims} \& \ref{satsims}), 
it is expected that such orbital perturbations will be more 
destructive to simulated satellite orbits than to those previously explored in Sect.~\ref{jupsect}. 
This notion is supported in Figs.~\ref{satevol}, \ref{satresults}, and \ref{finaleiap}
when compared to their counterparts in the Jupiter/IG case (Figs. \ref{jupevol}, \ref{jupresults}, 
and \ref{finalecall}) as the fraction 
of reconcilable Kronian satellites is systematically lower than in the Jupiter case.

The time evolution of $a_s$ and $e_s$ for five example satellites during a sample planetary 
encounter is shown in Fig. \ref{satevol}. This example is a particularly violent encounter with 
$b \approx 0.033$ AU and $v_{rel} \approx 5.87$ km/s as no satellite in this simulation has a resulting orbit 
that is reconcilable with Iapetus. One satellite, whose azimuthal position is 
only $4.6^{\circ}$ from $\theta(t_{enc})$, is ejected from Saturn by the encounter. 
The encounter timescale ($b/v_{rel} \sim 9.6$ days) 
is much less than the orbital period of Iapetus making the effect of the encounter behave 
approximately as an impulse. 

The average final orbital elements of all Kronian satellites after each unique 
encounter are depicted in Fig. \ref{satresults}. 
The black contours in Fig. \ref{satresults} are representative of the final orbital 
elements of the Jovian satellites from Fig. \ref{jupresults} to 
aid in visual comparison between the final 
orbits of Jovian satellites to Kronian satellites following similar close planetary 
encounters. This shows that the wide-separation of the 
simulated Kronian satellites makes them more susceptible to
excessive orbital perturbations. 

The event $RS$ in the Saturn/IG case is defined similarly to the Jupiter/IG case 
and is achieved by Kronian satellites whose final average $e_s \leq e_{Iapetus}$. 
As noted earlier, the eccentricity evolution of Iapetus since the solar system's 
instability phase is negligible \citep{castillo07} thus we assume that resulting 
orbits from our simulations will go largely unchanged to the present day. 
The boundary dividing $RS$ 
from non-$RS$ is depicted in Fig. \ref{satresults} as a dashed 
horizontal line. 

As noted in Sect.~\ref{jupresultssub}, no Jovian satellite becomes more eccentric than 0.4 
whereas a sizable fraction  
of Kronian satellites become equivalently or excessively excited including  
a subset of Kronian satellites whose 
final average $e_s$ is approximately unity. Furthermore, we find that $\sim 6$\% of 
sampled Kronian satellites are sufficiently excited to final $e_s>1$. 
The orbital elements of these ejected satellites are not included in Fig. \ref{satresults}. 
It is clear that the perturbations to the Kronian satellites resulting from close encounters, 
are on average much 
stronger and capable of stripping Iapetus-like satellites from the Kronian system. 
The large fraction of satellites 
perturbed beyond the orbit of Iapetus makes it statistically difficult for Saturn to have ejected 
an IG whilst retaining an Iapetus-like regular satellite. We formally estimate the likelihood of 
reconciling Iapetus' orbit in Sect.~\ref{prob}.

The effect of encounter properties on the resulting Kronian satellite orbits is shown in 
Fig. \ref{finaleiap}. For each encounter with 
a given impact parameter and relative planet velocity, 
the fraction of Kronian satellites whose final orbit is 
reconcilable with the current orbit of Iapetus is shown. Approximate contours of 10\% 
and 50\% fractions are over-plotted. The 10\% contour is computed identically to the 
contours in Fig. \ref{finalecall} (i.e. cubic interpolation). 
However, the 50\% contour lacks a sufficient number 
of points to perform a robust cubic interpolation. Therefore we opt for a linear fit 
in its place. It is clear that the region of the $(b,v_{rel})$ parameter space in which 
$\geq 50$\% of the Kronian satellites are reconcilable with Iapetus' orbit is 
very small compared to the Jovian satellites with only five unique ejections  
sampled there.

The effects of $b$ and $v_{rel}$ on resulting
satellite orbits are nearly identical to those shown in Fig. \ref{finalecall}.
One interesting difference which is unique to Saturn/IG encounters is that 
regardless of how close the 
encounter is, if the encounter is sufficiently long ($v_{rel} \lesssim 8$ km/s), 
then $< 10$\% of satellites will be reconcilable with Iapetus. 
This is in contrast to the Jovian 
satellite case where even the slowest encounters could retain a high fraction of 
Callisto-like satellites if the encounter's impact parameter is large. 
Another important feature to note is that no Saturn/IG encounter resulting in the ejection 
of the latter is guaranteed to preserve an Iapetus-like satellite. That is, even the least 
violent encounters leading to ejection are 
only capable of preserving a maximum of $\sim 70$\% of the in-situ Iapetus-like satellites.  

Fig. \ref{finaleiap} contains information on 
the likelihood that Iapetus survives an IG ejection by Saturn with a particular 
$b(t_{enc})$ and $v_{rel}(t_{enc})$ when the mutual inclination of the encounter 
is near zero. Similarly to the Jupiter/IG encounter case, we suggest that researchers 
simulating solar system formation scenarios can use Fig. \ref{finaleiap} to estimate 
whether or not a given Saturn/IG encounter is consistent with Iapetus' current orbit 
in the limit of uninclined planetary encounters. 

\section{Likelihood of Reconciling Satellite Orbits Following Ice Giant Ejections} \label{prob}

\subsection{Methodology} \label{probmeth}
The resemblance of the final average orbits of the Jovian and Kronian satellites 
(Figs. \ref{jupresults} and \ref{satresults}) to the current orbits of Callisto and Iapetus 
can, in principle, 
be used to compute the likelihood of a fifth giant planet getting ejected by either of the 
gas giant planets in the early solar system. If such an event were to have occurred, it must 
be consistent with the satellite orbits presently observed. 
The likelihood of an IG getting ejected by either gas giant 
requires the likelihood of the current orbits of Callisto or Iapetus  
being reconciled by simulated satellites after an IG ejection. 
A successful 
event in which the resulting orbit of a simulated Jovian (Kronian) satellite is reconcilable 
with Callisto's (Iapetus') current orbit is referred to as $RS$ for ``reconcilable satellite''.  

Given that an IG gets ejected in all simulations (event $IGE$; ``ice giant ejected''), 
for the $i^{\mathrm{th}}$ satellite in the 
$j^{\mathrm{th}}$ simulation, we record whether or not the event $RS$ is achieved 
by  

\begin{equation}
  p_{i,j}(RS|IGE) = \left\{
  \begin{array}{ll}
    1 & \quad \text{if $RS$} \\
    0 & \quad \text{if not $RS$}. \label{likely}
  \end{array} \right.\
\end{equation}

However, not all planet orbits in our ejection simulations are statistically relevant. 
Our adopted methodology for determining which encounter 
parameters result in an ejection is heavily biased towards initially high-eccentricity 
orbits of the IG ($e_I \lesssim 1$) 
which are not long-term stable and therefore are  
uncommon in nature. This bias naturally arises because it is easier to kick the IG to 
$e_I>1$ if $e_I$ is initially very close to unity. 
Therefore in each simulation $j$, 
each satellite's likelihood of $RS$ must be weighted by the likelihood of 
the IG having an initially bound orbit with $e_{I,j}$, where $e_{I,j}$ is the IG's 
initial eccentricity in the $j^{\mathrm{th}}$ simulation.

\begin{figure}
\epsscale{.9}
\includegraphics[scale=.43]{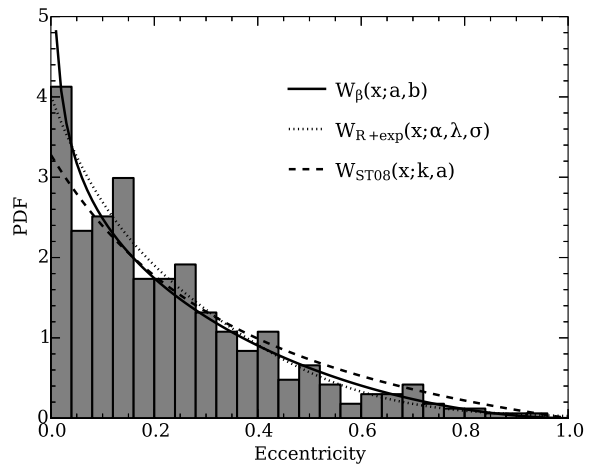}
\caption{Histogram of the planet eccentricities used to model the weighting function $W(e_{I,k})$ 
in Eqs. \ref{prob1} and \ref{norm}. The three analytical models used to fit the distribution are overplotted 
(\emph{solid line}: Beta PDF, \emph{dotted line}: Rayleigh PDF plus exponential, \emph{dashed line}: 
model from \cite{shen08}; see text). \label{dists}}
\end{figure}

The corresponding 
weighting function $W(e_{I,j})$ is modelled by the distribution of planet eccentricities 
observed in a statistically significant number of exoplanetary systems as the eccentricity 
distribution of solar system bodies is insufficient as it can   
only be derived in the limit of small number statistics. These data are recovered 
from the www.exoplanets.org database \citep{han14} 
and only include RV exoplanet detections with 
Doppler variation semi-amplitude $K/\sigma_K > 5$ (neglect low signal-to-noise observations). 
We find that our results do not sensitively 
depend on whether we focus on RV detections or the full catalogue of exoplanets with orbital solutions.
The distribution of planet eccentricities is shown in Fig. \ref{dists}. 
Due to variations in empirically derived eccentricity distributions   
we consider three proposed analytical forms of the distribution. 
Namely a Beta probability density function (PDF) \citep{kipping13b}, 
a Rayleigh PDF plus decaying exponential \citep{juric08,steffen10}, 
and the model from \cite{shen08}. 
We use a Levenberg-Marquardt least squares algorithm to fit for each PDF's unique parameters. 
A summary of the adopted distributions including fitted parameters is given in 
Table~\ref{tablepdfs} and each fitted weighting function is over-plotted in Fig.~\ref{dists}.   
Differences in $W(e_{I,j})$ from adopting three unique PDFs results in $\sim 6$\% variance 
among computed likelihoods.

\begin{table*}
\centering
\caption{Analytical fits to the planet eccentricity distribution. \label{tablepdfs}}
\begin{tabular}{ccccc}
  \hline
  \hline
  PDF Name & PDF & Parameter 1 & Parameter 2 & Parameter 3  \\  
  \hline
  Beta & $W_{\beta}(x;a,b) = \frac{1}{B(a,b)} x^{a-1} (1-x)^{b-1}$ & $0.786 \pm 0.055$ & $2.764 \pm 0.167$ & - \\
  Rayleigh $+$ exponential & $W_{\text{R+exp}}(x;\alpha,\lambda,\sigma) = \frac{x(1-\alpha)}{\sigma^2} \exp{(\frac{-x^2}{2\sigma^2})} + \alpha \lambda \exp{(-\lambda x)}$ & $0.782 \pm 0.301$ & $5.131 \pm 2.342$ & $0.266 \pm 0.061$ \\
  ST08 & $W_{\text{ST08}}(x;k,a) = \frac{1}{k}\Big(\frac{1}{(1+x)^a}-\frac{x}{2^a} \Big)$ & $0.305 \pm 0.012$ & $3.413 \pm 0.309$ & - \\
  \hline
\end{tabular}
\tablecomments{Parameter columns are written in the order in which they appear in the functional 
form $W(x;\dots)$, shown in the PDF column. 
We write the independent variable of planet eccentricity as $x$ to make the distinction 
between eccentricity and Euler's number $e \approx 2.7182818$.}
\end{table*}

While we take the known exoplanet orbital eccentricities as our nominal 
distribution, this sample is obviously biased by the requirement that systems be 
stable. If early on, the solar system’s ice giants moved on significantly more 
elliptical paths than the known exoplanets, it would be easier to eject planets 
while keeping Jupiter and Saturn’s satellites on nearly circular orbits, raising the 
likelihood of reconciling satellite orbits. 
However, this scenario would require a substantial amount of damping 
to subsequently recircularize the surviving planets’ orbits, and if eccentricities 
reached such high values, one might expect to also lose Uranus and Neptune. These 
considerations may be interesting directions for future work.

One must additionally weight $p_{i,j}$ by the simulation's impact parameter $b_j$, 
since wider encounters should occur more frequently.
Given our coplanar setup, the differential interaction cross-section 
for a given impact parameter scales linearly with $b_j$ rather than with $b_j^2$ as 
is true in the full 3D case. We find that our results do not sensitively depend on 
this distinction. We therefore assume $W'(b_j) \propto b_j$.

The distribution of $e_{I,j}$ is not explicitly prescribed and is instead determined from the 
encounter parameters ($b$,$v_{rel}$,$\theta$). The resulting distribution of $e_{I,j}$ 
is not sampled uniformly, unlike the distribution in $b$ or $\theta$, 
implying that the step size $\Delta e_I$ 
when integrating over IG eccentricities is not constant throughout the domain. 
Hence, simulations sample IG 
eccentricity bins of various widths which are taken into account when computing the likelihood 
function by calculating the width of IG eccentricity bins among the $N_{sim}$ simulations. 
In this way, the subspace 
of IG eccentricities which is over-sampled gets averaged over. Similar factors 
$\Delta b_j$ and $\Delta \theta_j$ are included but are constants, because the parameters 
are sampled uniformly, and therefore do not affect the resulting likelihood. 

Combining the aforementioned effects into the likelihood of obtaining a reconcilable 
satellite ($RS$) orbit given an IG ejection ($IGE$), we write  

\begin{multline}
  P(RS|IGE) = \\\frac{1}{\mu} \sum_{i=1}^{N} \sum_{j=1}^{N_{sim}} p_{i,j}(RS|IGE) W(e_{I,j}) W'(b_j) 
  \Delta e_{I,j} \Delta b_j \Delta \theta_j \label{prob1}
\end{multline}

\noindent where

\begin{equation}
  \mu = N \sum_{j=1}^{N_{sim}} W(e_{I,j}) W'(b_j) \Delta e_{I,j} \Delta b_j \Delta \theta_j, \label{norm}
\end{equation}

\noindent is the normalization factor of the weighted mean 
and we are careful to account for the fact that our uniform sampling of ($b$,$v_{rel}$,$\theta$)
leads to a non-uniform distribution of $e_{I,j}$. 

\subsection{Likelihoods}  \label{probresults}
From our $N_{sim}=278$ ejection simulations 
by Jupiter, each with $N$ satellites, 
we calculate that the likelihood of Jovian satellite orbits remaining consistent with the observed  
orbit of Callisto, is $\sim 42$\%. 
This value is the median of the results we obtained from adopting the 
three distinct eccentricity weighting functions discussed in Sect.~\ref{probmeth} (see 
Table \ref{tablepdfs}). 
The median absolute deviation among the likelihoods is small; $\lesssim 1$\%. 
 
By contrast, in our $N_{sim}=274$ ejection simulations by Saturn, the likelihood of Kronian 
satellite orbits remaining consistent with Iapetus' observed orbit is $\sim 1$\%, more than 
an order of magnitude less likely than for Callisto around Jupiter. 
The main reason for this wide likelihood disparity is that a given encounter will perturb 
Iapetus more strongly than Callisto because the former is less tightly bound to Saturn than 
the latter is to Jupiter. 
In this case, the median 
absolute deviation among the three eccentricity weighting functions is $\lesssim 0.6$\% 
which is comparable to the likelihood itself.  

\section{Discussion} \label{discuss}
To recapitulate, the above likelihoods assumed an IG ejection in a coplanar 
geometry, considering whether an initially circular Callisto (Iapetus) would have its 
orbital eccentricity excited beyond values reconcilable with its current orbit.  
We now consider these assumptions in turn, in order to interpret the results from our 
study and discuss possible implications of our work.

\subsection{Effect of Inclined Encounters} \label{inclined}
To limit the computational cost of our study, we have restricted our analysis to planetary 
encounters with no mutual inclination. However, if planetary eccentricities are excited
enough to permit orbit-crossing and ejections, one might expect comparably large orbital
inclinations. During an inclined encounter, some of the applied torque goes into realigning 
Callisto's (Iapetus') orbital plane  
so that, on average, the perturbations to $a_s$ and $e_s$ are reduced as some energy goes 
into increasing $i_s$. 

A preliminary analysis of inclined encounters with mutual inclination  
$i(t_{enc}) =5^{\circ}$ effectively revealed no change to the final satellite orbital elements 
compared to uninclined, but otherwise equivalent, encounters.
In a more heavily inclined test case with $i(t_{enc}) =45^{\circ}$,
we found that, on average, the
final $e_s$ were \emph{smaller} by $\gtrsim 0.01$ 
than in the uninclined case.

Therefore, by limiting our investigation to coplanar encounters, 
we are sampling the largest possible 
perturbations to $e_s$ without affecting $i_s$. The introduction of inclined encounters 
would thus raise the likelihoods quoted in Sect.~\ref{probresults}.
However, such a 3D case would additionally excite the satellite inclinations.  
As mentioned by \cite{deienno14}, these inclinations may provide more rigorous constraints 
on planetary encounters because even in cases where eccentricity damping is important, 
the tidal evolution of the inclinations is effectively null.
A generalization of this study would therefore consider the combined constraint from the 
satellites' orbital eccentricities and inclinations.  
However, the uncertainties in the likelihoods we compute in Sect.~\ref{probresults} are dominated by
the large uncertainties in the planetary orbital eccentricities (and inclinations) early 
in the solar system.
We therefore believe that our simplified analysis considering only the orbital eccentricities 
captures the
correct likelihoods at the approximate level allowable by our current state of knowledge.

As a rough check, in a fully 3D case, the differential cross-section for encounters of a 
given impact parameter would scale as $b_j^2$ rather than as $b_j$. 
Using the full data set from our coplanar 
simulations, but adopting this new scaling for $W'(b_j)$ in Eqs.\:\ref{prob1} and \ref{norm}, 
we find that 
our results do not change discernibly. Specifically, in the Jupiter case our results change 
from $\sim 42$\% to 54\% and from $\sim 1.1$\% to 1.3\% in the Saturn case.   
This is certainly within the errors of our uncertain knowledge of the planetary orbits' 
initial eccentricities and inclinations 
and therefore does not appreciably change the interpretation of our results.

\subsection{Interpretation of Results} \label{interp}
We conclude that Jupiter could plausibly have ejected an IG from the solar system.
Nearly half of the hypothetical set of ejections that were modelled in Sect.~\ref{probresults} keep Callisto
on an orbit which is reconcilable with the one we observe today. 
Put another way, 
we conclude that Callisto's orbit cannot meaningfully constrain whether or not Jupiter
ejected an additional IG in the early Solar System.  Nevertheless, this is an important test for
the fifth-giant-planet hypothesis to pass, thus providing more stringent evidence for its 
plausibility. 

Interpretation of our results in the Saturn case is more subtle.  
We showed in Sect.~\ref{probresults} that an initially circular Iapetus orbit
gets overly excited by a single ejection event $\sim 99\%$ of the time.  
This suggests that Saturn is not capable of ejecting an IG mass planet from the 
solar system.

But did Iapetus originally move on a circular path, as expected if it formed out of a 
circumplanetary disk? Starting with a circular orbit, 
all encounters act to raise the eccentricity. But if the
moon's initial path were instead elliptical, some ejection geometries could act to \emph{lower} 
the eccentricity, complicating the constraint that we nominally set. 
Perhaps one reason to doubt that Iapetus formed from a circum-kronian disk is that it 
is substantially inclined ($\approx 8^{\circ}$) to the local
equilibrium plane that one would expect it to follow. 

So what caused this aberrant inclination? \cite{hamilton13} recently suggested that 
Iapetus could indeed have formed on a circular orbit from a disk in its equilibrium plane, 
and subsequent collisions between Saturn's inner 
moons could have instead tilted the equilibrium plane for the exterior moons by the 
requisite amount.  In this case, our study implies that 
Saturn likely did not eject an IG from the solar system because Iapetus' relatively low 
eccentricity orbit cannot be reconciled with IG ejections by Saturn. 
On the other hand, Iapetus' inclination could be the signature of an ejection event itself.  
\cite{nesvorny14} recently studied such a 
scenario trying to simultaneously match Iapetus' current orbital eccentricity and inclination 
using simulations of the specific early solar system dynamical instabilities from 
NM12. They show that some cases are capable of sufficiently exciting Iapetus' 
orbital inclination whilst maintaining a low orbital eccentricity even for encounters as 
close as those considered in this study but not necessarily leading to ejection of the IG. 

In the Kronian case, therefore, the interpretation depends critically on the formation 
mechanism for Iapetus, which is currently unknown.  
If Iapetus' inclination is the result of collisions between inner moons \citep{hamilton13}, 
our results show that Saturn is unlikely to have ejected an IG from the early Solar System.  
By contrast, if Iapetus' orbit is fully explainable through
close planetary encounters, \cite{nesvorny14} showed that this requires many such 
close approaches which might be capable of ejecting the IG. 

\subsubsection{Single-Encounter Assumption}
We argue that the sole consideration of the final gas giant/IG encounter leading to ejection 
of the latter, is a sufficient measure of how close planetary encounters will modify regular 
satellite orbits. \cite{nesvorny14} showed for Iapetus, which is more susceptible to dynamical 
perturbations than Callisto, that numerous `soft' encounters prior to ejection have a fractional 
effect on satellite eccentricity compared to the eccentricity kick typically felt during the final 
encounter (Fig. \ref{jupresults} and \ref{satresults}). It should be noted that inclusion of numerous 
`soft' encounters prior to ejection would be detrimental to the computed likelihoods $P(RS|IGE)$ thus 
emphasizing that our results represent a conservative, best-case scenario.

\subsubsection{Close Encounters Without Ejection}
Due to the phase angle of an encounter being random, it is possible for a close encounter 
to have 
occurred at $0^{\circ} < \theta < 180^{\circ}$ and is subsequently not included in our likelihood calculation
as all  
ejection events occurred at $180^{\circ} \leq \theta \leq 360^{\circ}$ (see Figs.~\ref{jupsims} and 
\ref{satsims}). 
Any close encounter that does not lead to an ejection would still perturb regular 
satellite orbits, making their orbits prior to an IG ejection event, non-circular. 
The effect of an increased 
initial eccentricity reduces $P(RS|IGE)$, again making our calculation a conservative upper limit. 
However, because the frequency of close encounters with $0^{\circ} < \theta < 180^{\circ}$ is equivalent 
to the frequency of close encounters with $180^{\circ} \leq \theta \leq 360^{\circ}$, the effect on  
$P(RS|IGE)$ of close encounters without an ejection will only differ from our calculated values 
by a factor of order unity, such that we capture the correct order of magnitude on $P(RS|IGE)$.

\subsection{Additional Considerations} \label{future}
One sought after quantity relating to this work is the 
probability of an IG getting ejected by either gas giant given that the orbit of 
Callisto or Iapetus can be reconciled; $P(IGE|RS)$.
Using Bayes' theorem, this quantity could in principle be computed with knowledge of the 
likelihood functions calculated in this paper but also requires the independent probability of the IG 
getting ejected ($P(IGE)$; see NM12) and the normalization factor by the probability of 
a satellite exhibiting the current orbit of one of the wide-separation satellites ($P(RS)$). Despite 
the abundance of work done on the latter \citep[e.g.][]{canup02,estrada06,ward10,crida12,hamilton13,heller15}, 
precise constraints on $P(RS)$ are difficult to compute.
 
Also, a robust calculation of the likelihood of ejecting an IG from the solar system would require the 
event to be consistent with a vast number of constraints imposed by solar system bodies of which 
the orbital constraint $RS$ imposed by Callisto or Iapetus is just one. Therefore, 
such a calculation is 
not practical. However, consistency checks of IG ejections by a gas giant, such as those presented 
in this study, help to substantiate the proposed existence of a fifth giant 
planet. \emph{Based on our current understanding, which includes the main results of our study, 
there is little evidence demonstrating that an additional IG mass planet could not have existed 
in the early solar system}.  

\section{Summary} \label{summ}
Several studies trying to match the solar system's current orbital architecture argue 
for an early period of frequent planetary encounters 
\citep[e.g.][]{tsiganis05,brasser09,morbidelli09,levison11}.  
In addition, \cite{nesvorny11} 
found that adding a fifth giant planet to the solar system, which is subsequently ejected, 
better matches the 
current orbits of the remaining giant planets. 
In this paper, we therefore study whether such an ejection by either 
Jupiter or Saturn is reconcilable with the current observed orbits of their outermost 
regular satellites, Callisto and Iapetus. Our main conclusions are as follows:

\begin{itemize} \itemsep -2pt
\item The properties of planetary encounters (i.e. impact parameter, relative planet 
velocity, and encounter geometry) between Jupiter or Saturn and an 
unspecified ice giant, which are sufficiently violent to 
eject the latter, exhibit similar trends. 
\item The current (dynamically cold) 
orbit of the widest-separation Galilean satellite Callisto, 
has a significant likelihood ($\sim 42$\%) 
of being reconciled following the ejection of an ice giant  
planet from the solar system by Jupiter. 
\item Given the observed difficulty in 
reconciling the orbit of Iapetus with simulated Kronian satellites following an ejection event,  
the likelihood of Saturn ejecting 
an ice giant from the solar system is determined to be unlikely; likelihood $\sim 1$\%. 
\item However, we note the caveat that this interpretation is heavily dependent on the assumed 
formation scenario of Iapetus out of circum-kronian disk. Currently, the formation of Iapetus 
is largely uncertain.
\end{itemize}

We caution that these likelihoods should not be interpreted in an absolute sense. Rather, 
they are useful in showing that it is much easier for Jupiter to 
have ejected an ice giant than it is for Saturn. The evident 
plausibility of Jupiter being able to 
eject an IG thus supports the hypothesis of a fifth giant planet's existence in the 
early solar system.

\acknowledgments
RC would like to thank H. Rein, M. Van Kerkwijk, and R. Gomes for useful discussions 
and suggested 
improvements to the manuscript. RC also thanks D. Nesvorn{\'y} for numerical test cases 
of the solar system instabilities from NM12. 
Simulations in this paper made use of the collisional N-body code \texttt{REBOUND} 
which can be downloaded freely at http://github.com/hannorein/rebound. 
This research has made use of the Exoplanet Orbit Database 
and the Exoplanet Data Explorer at exoplanets.org.

\bibliographystyle{apj}
\bibliography{refs_solarsystem}

\end{document}